# Dichotomy in Low- and High-energy Band Renormalizations in Trilayer Nickelate La$_4$Ni$_3$O$_{10}$: a Comparison with Cuprates

X. Du[1,2†*], Y. L. Wang[3,4,5*], Y. D. Li[1], Y. T. Cao[6,7], M. X. Zhang[8], C. Y. Pei[8], J. M. Yang[2], W. X. Zhao[1], K. Y. Zhai[1], Z. K. Liu[8,9], Z. W. Li[6], J. K. Zhao[7], Z. T. Liu[10], D. W. Shen[11], Z. Li[12], Y. He[2], Y. L. Chen[8,9,13†], Y. P. Qi[9,10,14†], H. J. Guo[7†], and L. X. Yang[1,15†]

[1]*State Key Laboratory of Low Dimensional Quantum Physics, Department of Physics, Tsinghua University, Beijing 100084, China*

[2]*Department of Applied Physics, Yale University, New Haven, CT 06511, USA*

[3]*School of Emerging Technology, University of Science and Technology of China, Hefei 230026, China*

[4]*New Cornerstone Science Laboratory, University of Science and Technology of China, Hefei, 230026, China*

[5]*Hefei National Laboratory, University of Science and Technology of China, Hefei 230088, China*

[6]*School of Physical Science and Technology, Lanzhou University, Lanzhou 730000, China*

[7]*Songshan Lake Materials Laboratory, Dongguan, Guangdong 523808, China*

[8]*School of Physical Science and Technology, ShanghaiTech University, Shanghai 201210, China*

[9]*ShanghaiTech Laboratory for Topological Physics, Shanghai 200031, China*

[10]*Shanghai Synchrotron Radiation Facility, Shanghai Advanced Research Institute, Chinese Academy of Sciences, Shanghai 201210, China*

[11]*National Synchrotron Radiation Laboratory and School of Nuclear Science and Technology, University of Science and Technology of China, Hefei 230026, China*

[12]*Mork Family Department of Chemical Engineering and Materials Science, University of Southern California, Los Angeles, CA 90089, USA*

[13]*Department of Physics, Clarendon Laboratory, University of Oxford, Parks Road, Oxford OX1 3PU, UK*

[14]*Shanghai Key Laboratory of High-resolution Electron Microscopy, ShanghaiTech University, Shanghai 201210, China*

[15]*Frontier Science Center for Quantum Information, Beijing 100084, China*

*Emails: XD: xian.du@yale.edu; LXY: lxyang@tsinghua.edu.cn; YLC: yulin.chen@physics.ox.ac.uk; YPQ: qiyp@shanghaitech.edu.cn; HJG: hjguo@sslab.org.cn.*

**Abstract**

Band renormalizations comprise crucial insights for understanding the intricate roles of electron-boson coupling and electron correlation in emergent phenomena such as superconductivity. In this study, by combining high-resolution angle-resolved photoemission spectroscopy and theoretical calculations, we systematically investigate the electronic structure of the trilayer nickelate superconductor $La_4Ni_3O_{10}$ at ambient pressure. We reveal a dichotomy in the electronic band renormalizations of $La_4Ni_3O_{10}$ in comparison to cuprate superconductors. At a high energy scale of hundreds of meV, its band structure is strongly renormalized by electron correlation effect enhanced by Hund's coupling. The resultant waterfall-like dispersions resemble the high-energy kinks in cuprate superconductors. However, at low energy scales of tens of meV, the dispersive bands are nearly featureless and devoid of any resolvable electron-boson interactions, in drastic contrast to the low-energy kinks observed in cuprates and other correlated $3d$ transition-metal compounds. The dichotomic band renormalizations highlight the disparity between nickelate and cuprate superconductors and emphasize the importance of strong electron-correlation in the superconductivity of Ruddlesden-Popper phase nickelates.

**Introduction**

Since its discovery, high-temperature superconductivity (HTSC) in cuprates remains a key mystery in condensed matter physics [1,2]. To advance the understanding of this issue, the search for cuprate analogs with similar superconducting and normal-state properties has been intensely pursued. Strongly-correlated nickelates have emerged as promising candidates for high-temperature superconductivity, following the groundbreaking discovery of superconductivity in infinite-layer nickelates [3-5]. Notably, $Ni^+$ in these materials exhibits a $3d^9$-like configuration, analogous to $Cu^{2+}$ in cuprates [6,7]. However, their synthesis remains challenging, and superconductivity is observed only in thin films with relatively low transition temperature, even under applied pressure [3,5].

Recently, high-temperature superconductivity up to ~ 80 K was discovered in $La_{n+1}Ni_nO_{3n+1}$ in the Ruddlesden-Popper (RP) phase under pressure [8-13]. They exhibit extraordinary properties at ambient pressure, including orbital-dependent electronic correlation, non-Fermi liquid behavior, and the interplay with density-wave states [10,14-17]. Superconductivity at ambient pressure has also been recently stabilized in $(La,Pr)_3Ni_2O_7$ after ozone treatment [18,19]. Theoretical proposals for pairing mechanism mainly focus on the correlated electronic structure of the $d_{x^2-y^2}$ and $d_{z^2}$ orbitals [20-33]. Previous angle-resolved photoemission spectroscopy (ARPES) measurements have revealed the basic electronic structure and orbital-dependent electron correlation in $La_3Ni_2O_7$ [14,17,34] and $La_4Ni_3O_{10}$ [35]. However, systematic investigations at the mode-coupling and superconducting energy scales are still lacking partially due to insufficient experimental resolution and/or sample-quality-related issues.

Experimental exploration of the band structure of quantum materials is essential for understanding the microscopic interactions among different degrees of freedom. A prime example is the investigation of electron-boson coupling and electron correlation through band renormalizations. In this work, we use laser- and synchrotron-based high-resolution ARPES to investigate the low-energy mode coupling and high-energy band renormalization in $La_4Ni_3O_{10}$ single crystals. The dichotomy in the high- and low-energy band renormalizations, in comparison to cuprate superconductors, emphasizes the unusual nature of the electron excitations and highlights the important role of strong electron correlation rather than the electron-boson interaction in the superconducting transition of $La_4Ni_3O_{10}$.

**Results**

The similarity between nickelates and cuprates is reflected by their layered perovskite-related architectures. Figures 1(a)-1(d) compare the structure of Cu-O and Ni-O layers in trilayer cuprate $Bi_2Sr_2Ca_2Cu_3O_{10+\delta}$ (Bi-2223) and $La_4Ni_3O_{10}$. Both exhibit planar structures with square-type unit cell [Figs. 1(c) and 1(d)]. However, the Cu-O layers are isolated from each other, while the Ni-O layers are connected by the apical oxygen [Figs. 1(a) and 1(b)]. Moreover, the vertical axis of the Ni-O octahedrons tilts by an angle of about 8° with respect to the $c$ axis at ambient pressure. The application of pressure above 13 GPa aligns the octahedrons along the $c$ axis [36], which changes the crystal symmetry and induces superconductivity with a maximum $T_c$ of about 30 K [12]. Laue measurement [Fig. 1(e)] along the (001) plane confirms the symmetry and sample quality of the crystals in our experiments. The magnetic susceptibility decreases with temperature, followed by a sudden drop (enhancement) of the in-plane (out-of-plane) component near 132 K [Fig. 1(f)] due to the formation of the intertwined charge- and spin-density waves [37]. Correspondingly, specific heat shows a cusp near 132 K [Figs. 1(g) and 1(h)], in agreement with a previously reported density wave ordering on the Ni-O plane [37,38].

We present the basic band structure of $La_4Ni_3O_{10}$ in Fig. 2. The Fermi surface (FS) consists of three sheets: a circular pocket α, a large rounded-square pocket β, and a smaller square pocket δ [Fig. 2(a)], as indicated by the dashed red, blue, and orange lines in Fig. 2(b), respectively. The α pocket is better visualized by laser-ARPES measurement, showing an anisotropic distribution of the spectral weight along $\bar{\Gamma}\bar{S}$ and $\bar{\Gamma}\bar{X}$ [Fig. 2(c)]. The density-functional theory (DFT) calculated FS for the non-magnetic state [Fig. 2(d)] is consistent with the experimental measurement in general, except that an additional β′ pocket and a small oval pocket around Γ are predicted. Moreover, the experimentally observed δ pocket is missing in the calculation (see Supplemental Materials [39]).

Figures 2(e)-2(g) show the band dispersions along high-symmetry directions. The β band quickly disperses towards high energies, which is better visualized around the $\bar{\Gamma}'$ point in the second Brillouin zone [Fig. 2(e)]. The δ band shows much weaker intensity near $E_F$, forming the electron pocket δ around $\bar{\Gamma}'$ on the FS. Along $\bar{\Gamma}\bar{S}$ [Fig. 2(f)], there exists a flat band γ at about -30 meV near $\bar{\Gamma}'$ (see Supplemental Materials [39]), whose spectral weight forms patch-like feature on the FS [Fig. 2(a)]. Along $\bar{X}\bar{S}$ [Fig. 2(g)], the β band shows electron dispersion around $\bar{S}$ with the band bottom

near -100 meV. The overall electronic structure is in good agreement with the previous ARPES experiment [35], except that the δ band was not resolved before. Similar to the results in La$_3$Ni$_2$O$_7$ [14,17], the dispersive α/β bands and the flat γ band, contributed by the $d_{x^2-y^2}$ and $d_{z^2}$ orbitals respectively, exhibit an orbital-dependent correlation effect (see Supplemental Materials [39]).

After setting the basic multi-orbital character of the FS, we now turn to investigate the energy hierarchy of mode coupling on different FS sheets and along different high symmetry directions. Prominently, the experimental band structure is strongly renormalized to form waterfall-like dispersion at high binding energies. Such high-energy band renormalization is a main indicator of correlation-induced quasiparticle band destruction [40], and has been widely observed in cuprates, nickelates, and other correlated materials [41-46]. As compared in Figs. 3(a) and 3(b), the high-energy band renormalization occurs at a similar energy scale in the cuprate Bi$_2$Sr$_2$CaCu$_2$O$_{7+\delta}$ (Bi-2212) and La$_4$Ni$_3$O$_{10}$, which is higher than that in infinite-layer nickelates [47,48]. To quantify the energy scale of the band renormalization, we extract the energy deviation of the band dispersion from an assumed linear dispersion [dashed white line in Fig. 3(b)] as the real part of the electron self-energy (ReΣ). The peak in ReΣ reveals the high-energy band renormalization near -220 meV [Fig. 3(c)]. The imaginary part of the electron self-energy (ImΣ) [Fig. 3(d)] gradually increases with the binding energy, showing a change of the slope near -220 meV. Both ReΣ and ImΣ of La$_4$Ni$_3$O$_{10}$ are in good accordance with those in cuprates [41,42].

The observed high-energy band renormalization cannot be understood by boson mode coupling since the energy of phonons and magnetic excitations are much lower than the high-energy waterfall position in RP-phase nickelates [49-51]. Rather, the waterfall-like structure is believed to appear as strong Coulomb interaction $U$ splits quasiparticle bands into lower- and upper-Hubbard bands [52]. In multi-orbital systems like nickelates, the Hund's coupling $J_H$ further modulates the electron correlation by inducing an orbital-dependent electron correlation effect. Figs. 3(e) and 3(f) compare the DMFT calculations of La$_4$Ni$_3$O$_{10}$ with different $J_H$ at Hubbard $U$ = 4 eV. The Hund's coupling greatly broadens the spectra at high binding energies [53]. In particular, for $J_H$ = 1 eV, the band bottoms at $\bar{\Gamma}$ and $\bar{S}$ are smeared out [Fig. 3(g)], leaving a waterfall-like dispersion, consistent with our experiment in Figs. 2(e)-2(g). Correspondingly, ImΣ of Ni $e_g$ orbitals rises and tends to saturate at high binding energy [Fig. 3(h)], showing agreement with our experiments. Therefore, we conclude that the high-energy band renormalization in La$_4$Ni$_3$O$_{10}$ originates primarily from the strong electron correlation effect enhanced by Hund's coupling.

After revealing the high-energy band renormalizations, a natural question then arises as to whether RP-phase nickelates exhibit low-energy kinks induced by electron-boson coupling as in cuprates [54-59]. Figures 4(a) and 4(b) compare the band dispersions of Bi-2212 and La$_4$Ni$_3$O$_{10}$ along the direction oriented 45° relative to the in-plane Cu-O/Ni-O bond. While the nodal band in Bi-2212 exhibits a clear kink structure near -70 meV [Fig. 4(a)], the β band in La$_4$Ni$_3$O$_{10}$ shows a linear dispersion to about -200 meV without evident kink [Fig. 4(b)]. Similarly, the β band at $\bar{S}$ [Fig. 4(c)] fits nicely to the DFT-calculated result, likewise without any kink structure (see Supplemental Materials [39]). The distinction in the low-energy band renormalization of the two compounds is demonstrated by the comparison of ReΣ in Fig. 4(d). The ReΣ shows a prominent peak near -70 meV in Bi-2212 corresponding to the planar Cu-O bond half breathing phonon [57], while it is nearly featureless with |ReΣ| well below 5 meV in La$_4$Ni$_3$O$_{10}$, despite the same phonon excitation expected from the similar planar metal-oxygen structures.

The notable absence of low-energy kink in La$_4$Ni$_3$O$_{10}$ is further evidenced by our temperature-dependent laser-ARPES measurements along $\bar{\Gamma}\bar{S}$ [Figs. 4(e)-4(g)]. The α band dispersion only shows a closing density-wave gap near $E_F$ [Fig. 4(f)] [60]. To simulate the effect of gap opening on quasiparticle dispersion, we impose an energy gap in the dispersion relation by using $E_\mathbf{k}(T) = \sqrt{\epsilon_\mathbf{k}^2 + \Delta(T)^2}$, where $\epsilon_\mathbf{k}$ is the dispersion at high temperature (solid red curve) and $\Delta(T)$ is the temperature-dependent density-wave gap, with $\Delta$ ($T$ = 0 K) = 25 meV and transition temperature $T_{\text{dw}}$ = 132 K. The simulated curve (dashed black curve) excellently reproduces the band dispersion at 25 K (solid blue curve). By subtracting the simulated dispersions $E_\mathbf{k}(T)$ from the measured dispersions at different temperatures, we confirm the vanishingly small change of the α band with temperature [Fig. 4(g)]. Moreover, we emphasize that ImΣ of the α band also shows minor temperature dependence without any visible energy anomaly (see Supplemental Materials [39]). These observations distinctly diverge from the temperature evolution of the band renormalizations induced by electron-phonon coupling in cuprates [55,59] and also deviate from the expectation of Migdal-Eliashberg theory [61,62] (see Supplemental Materials [39]).

**Discussion**

The absence of dispersion anomaly in low-energy band dispersions indicates weak electron-boson interaction in $La_4Ni_3O_{10}$, consistent with recent ultrafast reflectivity measurements [51] and theoretical predictions [63,64]. Nickelates in RP-phase such as $La_4Ni_3O_{10}$ consist of Ni-O planar structure and partially-filled $d_{x^2-y^2}$ orbital. The $d_{x^2-y^2}$ orbital directly hybridizes with the in-plane O $p_x/p_y$ orbitals (see Supplemental Materials [39]) and is in principle prone to couple with in-plane lattice vibrations [65]. As the most coupled phonon modes involve oxygens [62,64], the distinction between nickelates and cuprates may arise from different charge-transfer energy $\Delta_{CT}$ between Ni $3d$ and O $2p$ orbitals. Since $\Delta_{CT}$ in nickelates is twice as strong as in cuprates [21,66], $3d$-$2p$ mixing is reduced in nickelates. As a result, the electron-phonon coupling in nickelates is generally expected to be weaker than in the cuprates. In addition, any electron coupling to magnetic excitations is restricted within tens of meV range in RP nickelates [49], which is also clearly absent in the current experiment. To our best knowledge, our high-resolution data place by far the most stringent upper bound on the low-energy mode coupling strengths in multilayer nickelates [17,35,48,67].

Finally, although the dichotomy of low- and high-energy band renormalizations is revealed at ambient pressure, similar physics should persist in the superconducting phase. On the one hand, it has been proposed that strong electron correlation modulated by Hund's coupling is essential in the superconductivity of RP-phase nickelates at high pressure [68], for which band renormalizations at high-energy scales are theoretically predicted [69,70]. On the other hand, weak electron-phonon coupling is predicted for the superconducting phase in the sister compounds, infinite-layer nickelate [71] and bilayer RP-phase nickelate $La_3Ni_2O_7$ [63,64,72]. Also, similar dichotomic band renormalizations exist in infinite-layer nickelates [47,48], suggesting that electron correlation rather than the electron-boson interaction is likely at play in the superconducting phase as well. Therefore, unlike the contention of whether low-energy boson-mediated pairing or high-energy correlation-induced pairing should be responsible in cuprates, our results unambiguously point to the necessity to directly consider strong correlation effects behind the pairing mechanism in nickelate superconductors.

In conclusion, strong electron correlation induces waterfall-like features in $La_4Ni_3O_{10}$ resembling the high energy kink in cuprates, while the low-energy kink originated from electron-boson coupling is absent at low energies. This dichotomy makes RP-phase nickelates unique in the 3$d$ transition metal compounds and highlights the electron correlation in understanding the unconventional superconductivity of nickelates.

**Acknowledgement**

This work is funded by the National Key R&D Program of China (Grant No. 2022YFA1403201 and Grant No. 2022YFA1403100), the National Natural Science Foundation of China (No. 12274251, 12004270, 52272265, and 12174365), and the Guangdong Basic and Applied Basic Research Foundation (Grant No. 2022B1515120020). L.X.Y. acknowledges support from the Tsinghua University Initiative Scientific Research Program and the Fund of Science and Technology on Surface Physics and Chemistry Laboratory (No. XKFZ202102). We thank the MAX IV Laboratory for the time on beamline Bloch under Proposal 20230668, and the SSRF for the time on beamline 03U under Proposal No. 2023-SSRF-PT-502880. Y.L.W. acknowledges support from the Innovation Program for Quantum Science and Technology (No. 2021ZD0302800) and the New Cornerstone Science Foundation.

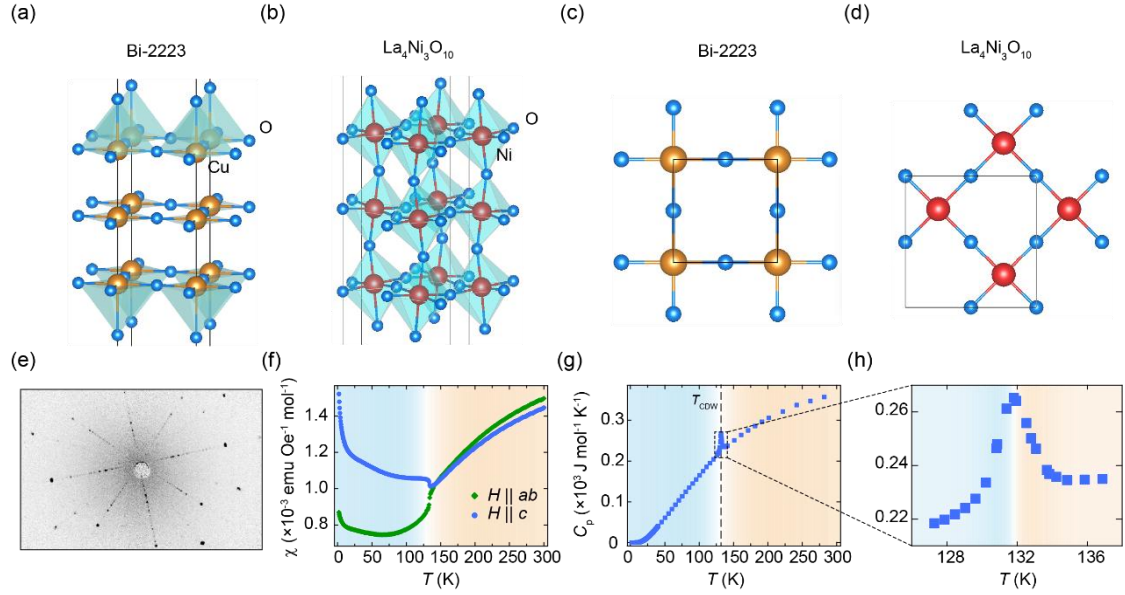

FIG. 1. Basic properties of $La_4Ni_3O_{10}$. (a, b) Crystal structure of $Bi_2Sr_2Ca_2Cu_3O_{10}$ (Bi-2223) (a) and $La_4Ni_3O_{10}$ (b) at ambient pressure. (c, d) Projection view of $CuO_2$ (c) and $NiO_2$ (d) planes. Only Cu/O and Ni/O atoms are shown. (e) Laue pattern of $La_4Ni_3O_{10}$ along (001) plane. (f, g) Magnetic susceptibility (f) and specific heat (g) as a function of temperature. (h) Magnification of the cusp near 132 K.

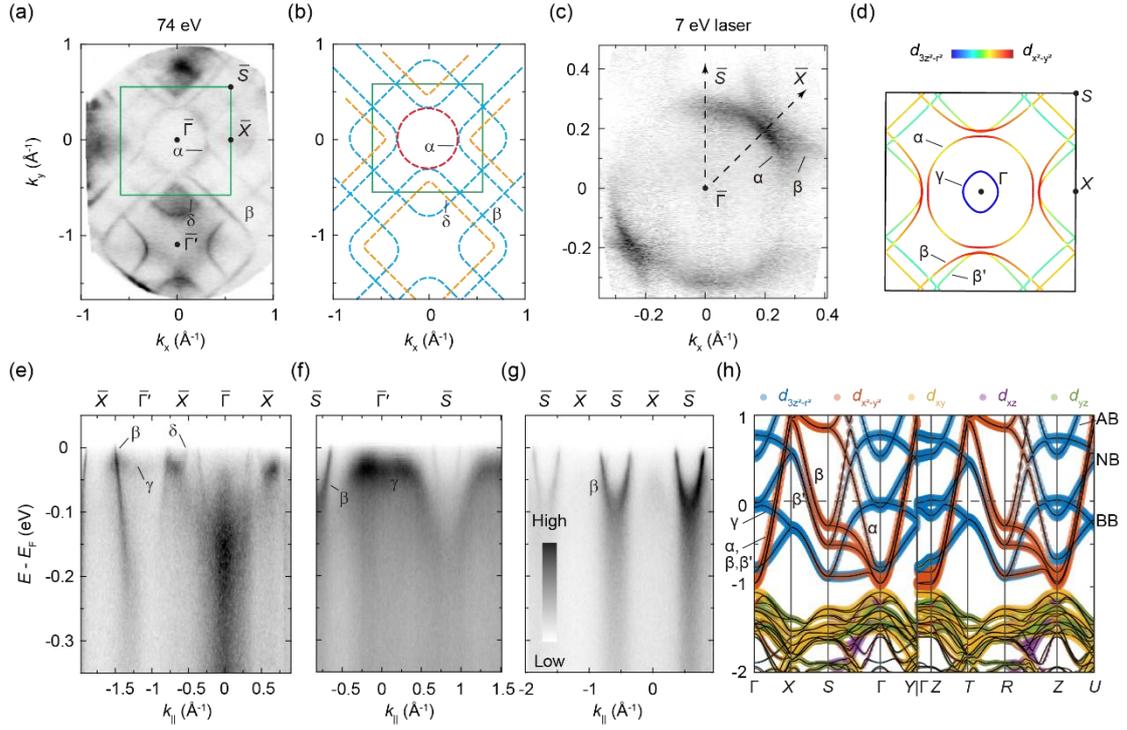

FIG. 2. Electronic structure of La$_4$Ni$_3$O$_{10}$. (a) Experimental Fermi surface (FS) of La$_4$Ni$_3$O$_{10}$ integrated within an energy window of ± 20 meV around the Fermi level ($E_F$). Data were collected with photon energy of 74 eV. (b) Sketch of the FS as a guide of eyes for the experimental result in (a). The surface Brillouin zones (BZ) are overlaid as green rectangles in (a, b). (c) FS measured with 7-eV laser, integrated within an energy window of ± 20 meV around $E_F$. (d) Density-functional-theory calculated FS. (e-g) Band dispersions along the high-symmetry directions of $\bar{\Gamma}\bar{X}$ (e), $\bar{\Gamma}\bar{S}$ (f), and $\bar{X}\bar{S}$ (g) measured at the photon energy of 98 eV (e, g) and 74 eV (f). (h) Calculated band structure projected onto different Ni 3$d$ orbitals. BB, bonding band. NB, non-bonding band. AB, anti-bonding band. Data in (c) were collected with linear-vertically (LV) polarized photons at 80 K. All other data were collected with linear-horizontally (LH) polarized photons at 20 K.

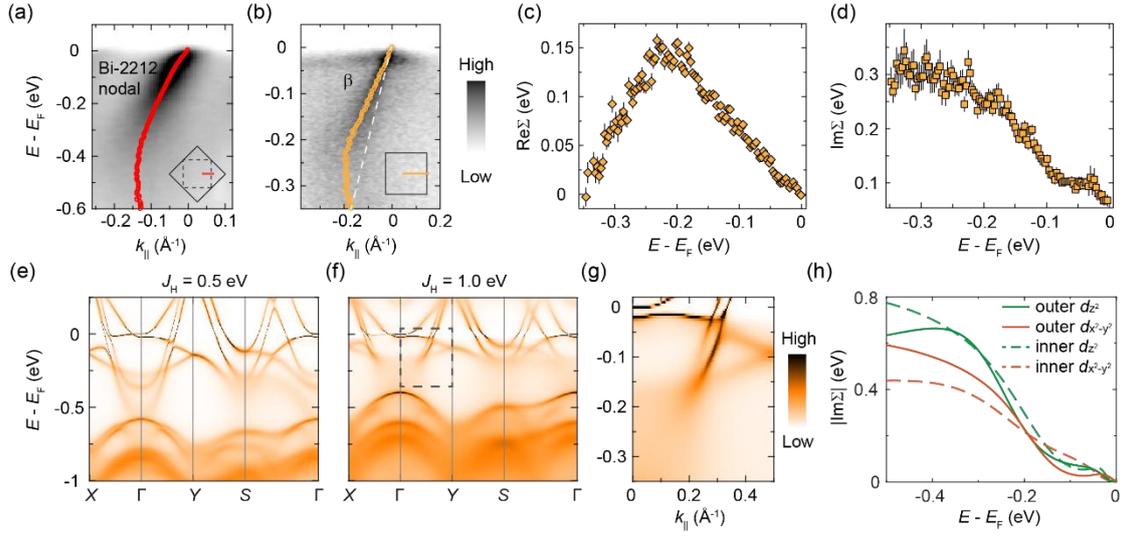

FIG. 3. High-energy band renormalization. (a, b) Spectrum of $Bi_2Sr_2CaCu_2O_{8+\delta}$ (Bi-2212) (a) and $La_4Ni_3O_{10}$ (b), with band dispersion extracted by fitting momentum-distribution curves (MDCs) overlaid. The dashed white line in (b) indicates an assumed linear dispersion. (c) The deviation of the β band from the linear dispersion in (b), which resembles the real part of electron self-energy. (d) Imaginary part of the electron self-energy extracted from the MDC-fitting in (b). Data in (a) were measured with LH-polarized 55 eV photons at 28 K. Data in (b-d) were measured with LH-polarized 98 eV photons at 20 K. (e, f) Dynamical-mean-field-theory (DMFT) calculated spectral function of $La_4Ni_3O_{10}$ for Hund's coupling $J_H = 0.5$ eV (e) and $J_H = 1.0$ eV (f). (g) Zoom-in plot in the dashed grey rectangle in (f) showing a waterfall-like structure. (h) Calculated orbital- and site-dependent self-energy with $J_H = 1.0$ eV. Hubbard $U = 4$ eV was applied in the calculations.

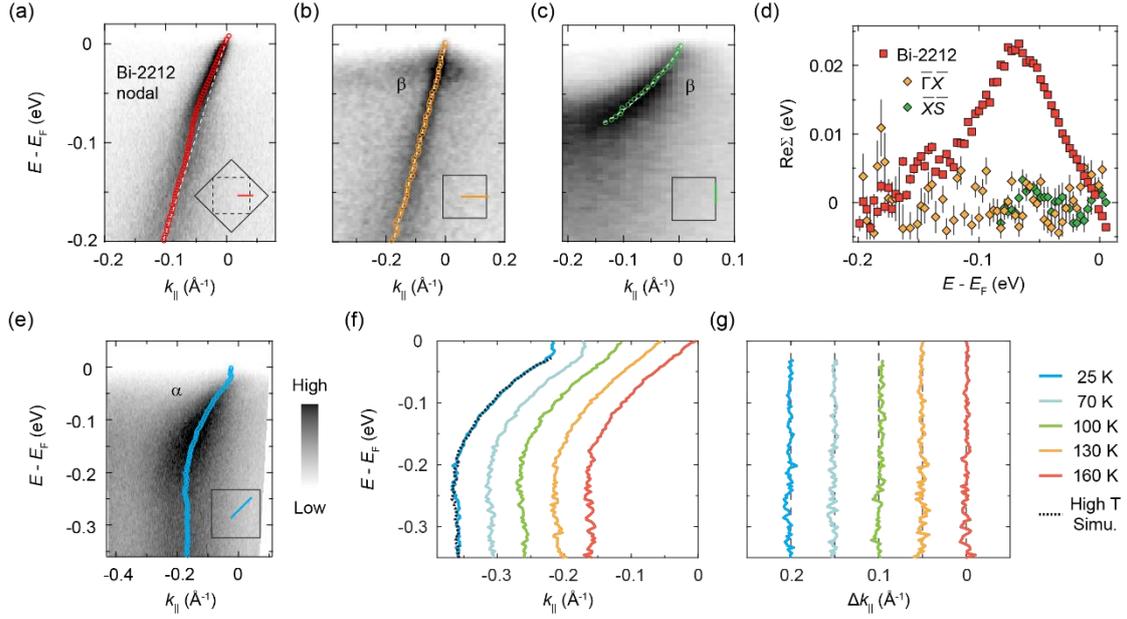

FIG. 4. Low-energy band renormalization. (a) ARPES-measured band dispersion of Bi-2212 along the nodal direction. (b, c) Measured band dispersion of La$_4$Ni$_3$O$_{10}$ along the $\bar{\Gamma}\bar{X}$ (b) and $\bar{X}\bar{S}$ (c) directions. The colored circles represent the band dispersion extracted by MDC-fitting and the dashed white lines represent the bare band. (d) Real part of the electron self-energy extracted from spectra in (a-c). (e) Measured dispersion of the α band along the $\bar{\Gamma}\bar{S}$ direction. The blue circles represent fitted band dispersion. (f) Temperature evolution of the α band. Dashed black line overlaid on the data at 25 K is the simulated curve with a gap of 25 meV enforced in the high-temperature data (average of data measured at 145 K, 160 K, and 175 K). (g) Difference curves of band dispersion with respect to the high temperature data with enforced temperature-dependent gap. Curves in (f) and (g) are horizontally shifted by steps of 0.1 Å$^{-1}$ and 0.05 Å$^{-1}$ with respect to the 160 K curve, respectively. Data in (a) were collected with LH-polarized 7 eV photons at 20 K. Data in (b, c) were collected with LH-polarized 98 eV photons at 20 K. Data in (e-g) were collected with LV-polarized 7 eV photons at 25 K (e) and indicated temperatures (f, g).